\documentclass[review]{elsarticle}
\usepackage{multirow}
\usepackage{color}
\usepackage{amsmath}
\usepackage{amssymb}
\usepackage{graphicx}
\usepackage{lineno,hyperref}
\modulolinenumbers[5]

\journal{}

\begin{document}

\begin{frontmatter}

\title{Designing rare-earth free permanent magnets in Heusler alloys via interstitial doping}


\author[mymainaddress]{Qiang~Gao}

\author[mymainaddress]{Ingo~Opahle}

\author[mymainaddress,secondad]{Oliver~Gutfleisch}

\author[mymainaddress]{Hongbin~Zhang}
\cortext[mycorrespondingauthor]{Corresponding author}
\ead{hzhang@tmm.tu-darmstadt.de}

\address[mymainaddress]{Institut f\"ur Materialwissenschaft, Technische Universit\"at Darmstadt, 64287, Darmstadt, Germany}

\address[secondad]{Fraunhofer-Research Institution Materials Recycling and Resource Strategies IWKS, 63457, Hanau, Germany}

\begin{abstract}
Based on high-throughput density functional theory calculations, we investigated the effects of light interstitial H, B, C, and N atoms 
on the magnetic  properties of cubic Heusler alloys, with the aim to design new rare-earth free permanent magnets. 
It is observed that the interstitial atoms induce significant tetragonal distortions, leading to 32 candidates with large ($>$ 0.4 MJ/m$^3$) uniaxial magneto-crystalline anisotropy energies (MAEs) and 10 cases with large in-plane MAEs. Detailed analysis following the the perturbation theory and chemical bonding reveals the strong MAE  originates from the local crystalline distortions and thus the changes of the chemical bonding around the interstitials.
This provides a valuable way to tailor the MAEs to obtain competitive permanent magnets, filling the gap between high performance Sm-Co/Nd-Fe-B and widely used ferrite/AlNiCo materials.
\end{abstract}

\begin{keyword}
Permanent magnets \sep Interstitial \sep Tetragonal distortion \sep Magneto-crystalline anisotropy energy
\MSC[2010] 00-01\sep  99-00
\end{keyword}

\end{frontmatter}

\section{Introduction}

Permanent magnets are of great technical importance for many key technologies such as electric vehicles, wind turbines, and automatisation and robotics to name only a few~\cite{gutfleisch_magnetic_2011}.
Looking at the intrinsic magnetic properties, such materials demand a large 
magneto-crystalline anisotropy energy (MAE), a sizable saturation magnetization, and a high Curie temperature. 
The MAE originates from the spin-orbit coupling (SOC) and sets an upper limit for the microstructure dependent coercivity of permanent magnets.
At present, rare-earth magnets based on Sm-Co (MAE: 17.0 MJ/m$^3$, Magnetization (M$_s$): 910 kA/m) and Nd-Fe-B (MAE: 5.0 MJ/m$^3$, M$_s$: 720 kA/m) are prototypes of high performance permanent magnets, 
 with a substantial cost and performance  gap to other classes of commercially available permanent magnets such as AlNiCo (MAE: 0.04 MJ/m$^3$, M$_s$: 50 kA/m) and ferrites (MAE: 0.03 MJ/m$^3$, M$_s$: 125 kA/m)~\cite{coey_permanent_2012}. 
Thus, there is a great interest to develop novel permanent magnets so that the full spectra of applications can be achieved, 
ideally without critical elements such as rare-earth elements~\cite{jpd_current, Skokov-2018}.

An enlightening idea was proposed to achieve giant MAE in tetragonally distorted FeCo alloys~\cite{prl-feco-1},  
where both the tetragonal distortion and fine tuning of the number of electrons by alloying are crucial for the enhanced MAE. 
Follow-up experimental studies on FeCo alloys deposited on various substrates confirmed the theoretical prediction~\cite{prl-feco-pt}.  
Nevertheless, due to the strong tendency for the FeCo alloys to relax, it is difficult to maintain the tetragonal distortion 
induced by the underlying substrates for thin films thicker than 2 nm~\cite{prl-feco-pt, apl-91-262512-2007, prb-80-064415-2009}. 
Recently, following the prediction based on DFT calculations~\cite{jap-116-213901-2014, j-physics-condensed-matter-27-476002-2015},
systematic studies have been performed on FeCo+X (X= C and B), where spontaneous tetragonal distortions
with c/a=1.04 can be induced by a few atomic percent interstitial  doping of C or B atoms occupying the octahedral interstitial sites. 
The resulting MAE can be as large as
0.5 MJ/m$^3$ with B concentration up to 4 at\%, where the tetragonal strain reaches 5\%.  
For Fe$_{0.38}$Co$_{0.62}$, a large interstitial concentration of 9.6 at\% B was achieved.~\cite{j-physics-condensed-matter-27-476002-2015}
The effect of light interstitials on the magnetic properties of body-centered cubic (BCC) iron has also been well studied.
$\alpha$-Fe with 12.5 at\% content of nitrogen interstitial has been grown by sputtering on the MgO (100) substrates, leading 
to about 10\% tetragonal distortion and significant enhancement of magnetization and MAE~\cite{fe-n-k}. 
First-principle calculations and experimental results show that Fe  
with nitrogen interstitial has 
sizable MAE, favoring perpendicular magnetization~\cite{fe-n-k}. 
Using the molecular beam epitaxy, boron has been incorporated into bcc Fe as interstitial dopants, which give rise to tetragonal distortions but the resulting MAE still favors in-plane magnetization due to tendency for B atoms to be agglomerated~\cite{fe-b}, where the interstitial content of B atoms can be as high as 14 at\%.

Considering only the crystal structure, the austenite phase of Heusler alloys with the conventional cubic cell can be regarded as a $2\times2\times2$ supercell of the bcc lattice.
In this regard, light interstitials such as H, B, C, and N  can also be promising to induce significant tetragonal distortions and thus substantial MAE to Heusler alloys, like the FeCo alloys and bcc Fe.
It is noted that the Heusler alloys in the tetragonal martensitic phase do show significant MAE.
For instance, among 286  
Heusler compounds, a systematic high throughput (HTP) screening 
suggests 19 potential tetragonal systems with large out-of-plane MAE (as large as 0.9 MJ/m$^3$)~\cite{fprmheusler}.
Matsushita  {\it et al.} found 15 Heusler compounds have tetragonal distortions of which the MAEs ranges from -12 MJ/m$^3$ to 5.19 MJ/m$^3$~\cite{jpd-Matsushita}.
Focus on Ni based full Heusler compounds, Herper $et. al.$~\cite{ni-based} found tetragonal Ni$_2$FeGe has an MAE of 0.95 MJ/m$^3$, 
which can be further increased to 1 to 2 MJ/m$^3$ by non-magnetic doping.  
Furthermore, imposing strain by proper substrates is helpful to engineer a large MAE out of the cubic Heusler alloys.
It is found that the out-of-plane MAE of epitaxial Co$_2$MnGa (001) films can be remarkably enhanced from 0.11 MJ/m$^3$ to 0.33  MJ/m$^3$ by changing the substrate from  ErAs/InGaAs/InP  to ScErAs/GaAs~\cite{jmm-Michael-286-340-2005}.
Lastly, previous experiments have already demonstrated that interstitials can be incorporated into Heusler alloys, leading to enhanced mechanical stability and magnetocaloric effect~\cite{Scripta-Materialia-75-15-2014, apl-100-192402-2012}.
For  Ni$_{43}$Mn$_{46}$Sn$_{11}$C$_x$, when the interstitial content x is increased from 0 to 8 the martensitic phase transformation temperature is increased from  196 to 249 K,   while a remarkable increase of MAE is observed when x is increased from 0 to 2~\cite{Scripta-Materialia-75-15-2014}. 
Due to large loss of manganese in content of x=8, there is even a distortion of crystal structure from Hg$_2$CuTi-type to the Cu$_2$MnAl-type~\cite{Scripta-Materialia-75-15-2014}.
Similar effect has also been observed in Ni$_{50}$Mn$_{34.8}$In$_{14.2}$, Ni$_{43}$Mn$_{46}$Sn$_{11}$ and Ni$_\text{50}$Mn$_\text{38}$Sb$_\text{12}$ doped with B interstitial~\cite{apl-100-192402-2012, apl-92-102503-2008, MSE-B-176-1322-1325-2011}.

In this work, focusing on developing rare-earth free permanent magnets, we have performed high-throughput first-principles calculations to investigate the
effects of light interstitials ({\it e.g.}, H, B, C, and N) on cubic Heusler alloys. 
After identifying the most favorable site preference of the interstitial atoms,  the MAE of compounds with negative formation energy
was evaluated to select the most promising candidates. Apart from thermodynamically stable  criteria, the disorder effect should also be considered, which is however beyond the scope of the present paper and saved for future study.
We observed that the induced MAE can be as large as 2.4~MJ/m$^3$, and there are 32  systems with a sizable out-of-plane MAE (> 0.4~MJ/m$^3$). 
Detailed analysis based on the Bain path and the atom-resolved MAE reveal that not only the global tetragonal distortion
but also the associated local chemical bonding are crucial for the interstitial induced magnetic anisotropy.



\section{Computational details}
\label{comdetail}

Starting with 128 full 
Heusler alloys with space group Fm$\bar{\text{3}}$m including at least one of magnetic atoms Cr, Mn, Fe, Co, and Ni from the Inorganic Crystal Structure Database (ICSD)~\cite{icsd} (cf. 
Table~\ref{tables1} in Appendix~\ref{appendx}),
we performed density functional theory (DFT) calculations firstly to identify the energetically most favored interstitial sites for H, B, C, and N atoms.
There are four types of interstitial sites based on the symmetries, as shown in Fig~\ref{fig:1}(a). The DFT calculations are managed with our in-house developed high-throughput environment (HTE)~\cite{hte1,hte2}, 
using both the Vienna ab initio Simulation Package (VASP)~\cite{vasp1,vasp2} and full-potential local-orbital (FPLO)~\cite{fplo1,fplo2} codes.
The structure optimization is performed in a two step manner. 
Firstly, ultrasoft pseudopotentials (US-PP)~\cite{chaoyan} are used in combination with the PW91~\cite{vasp3}  exchange correlation functional, 
where the cutoff energy for the plane wave basis is set to 250 eV and and a k-mesh density of 30\AA$^{-1}$. 
Secondly, the structure is relaxed  using the projector augmented plane wave (PAW) method with the exchange-correlation functional under 
the generalized gradient approximation (GGA) parameterized by Perdew, Burke, and Ernzerhof (PBE)~\cite{vasp4} with increasing plane 
wave expansion as 350 eV and $k$-mesh density as 40 \AA$^{-1}$ to achieve good convergence. 
After obtaining the energy lowest configuration, the MAEs of candidates with negative formation energy are calculated by using FPLO with a $k$-mesh density of 120 \AA$^{-1}$ to guarantee 
fine convergence.  For the MAE calculations of Ni$_2$FeGa with C interstitial, the resulting $k-$mesh is set as $24\times 24\times 17$.  
The bonding analysis is done in terms of the crystal orbital Hamilton population (COHP) evaluated using the LOBSTER code~\cite{cohp}.


\section{Results and discussions}

\begin{figure}
\centering

\includegraphics[width=6cm]{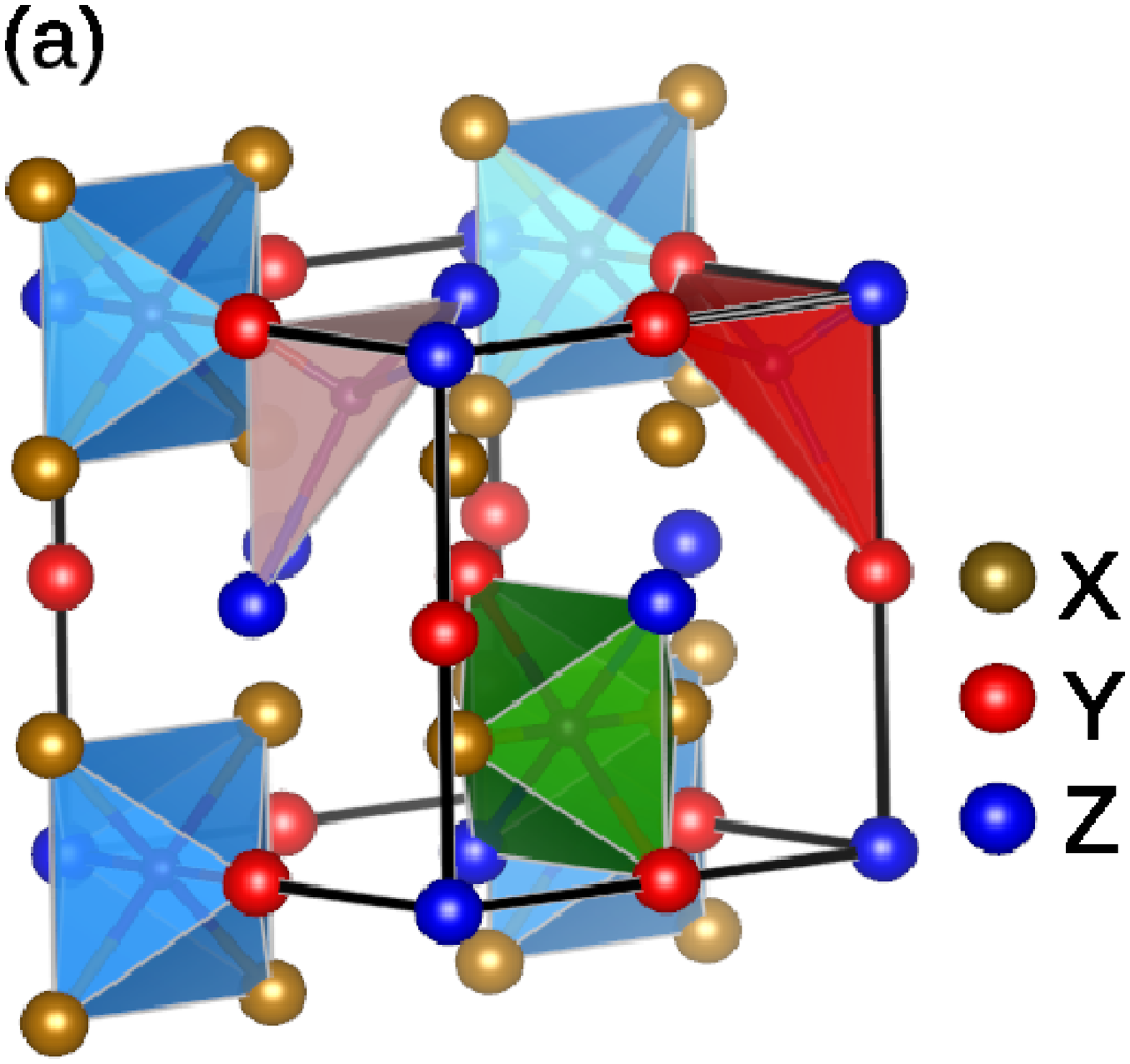}
\includegraphics[width=6cm]{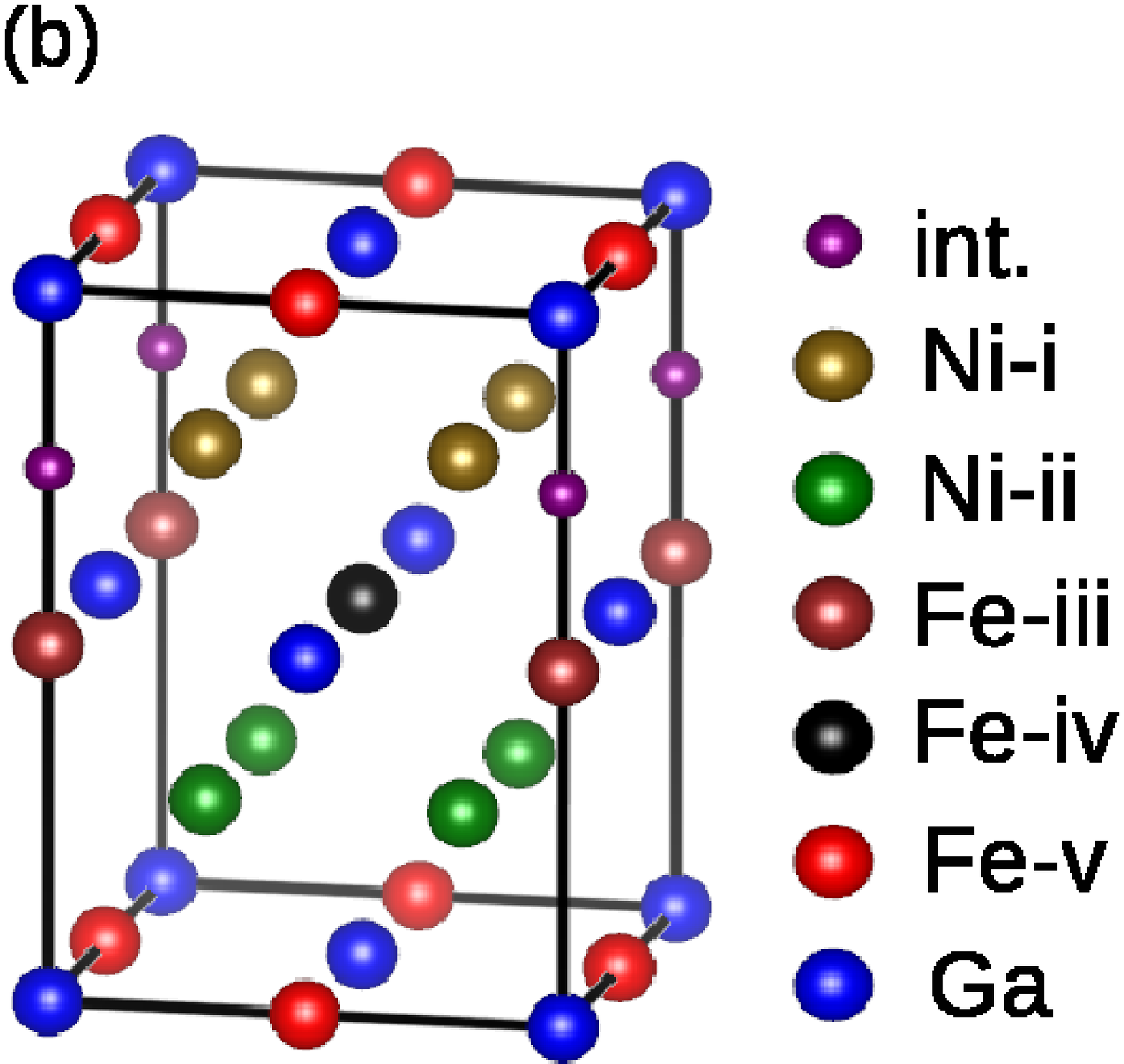}

\caption{}

\label{fig:1}
\end{figure}

As shown in Fig.~\ref{fig:1}(a), the systems we considered correspond to doping 6.25 at\% interstitial atoms (I) into the full Heusler alloys (X$_2$YZ), leading to a general chemical formula X$_2$YZI$_{1/4}$. 
This is in accordance with the typical doping concentrations experimentally accomplishable, $e.g.$, 12.5 at\% content of N in Fe and 9.6 at\% of B in Fe$_{0.38}$Co$_{0.62}$. ~\cite{j-physics-condensed-matter-27-476002-2015, fe-n-k}
Like Fe-Co alloys, we find light interstitials can indeed cause stable tetragonal distortion to cubic full Heusler alloys, which is quantizated by the c/a ratio between the c-axis and in-plane lattice constants.
As shown in Table~\ref{table1}, with N interstitials, Fe$_2$NiAl has the a tetragonal distortion as large as c/a=1.57.
Such strong tetragonal distortions prevail in the other Heuslers with the other types of interstitial atoms, which break the cubic symmetry and hence lead to possible significant MAE.
From the theoretical point of view, the MAE is defined as the total energy difference between the magnetization directions parallel to [100] (in-plane) and [001] (out-of-plane) 
directions as
\begin{equation}
\label{eqm}
\text{MAE}=E_{[100]}-E_{[001]}
\end{equation}
where $E_\alpha$ is the total energy when magnetization direction is parallel to 
${\alpha}$. 
When the MAE value is positive (negative),  the spontaneous magnetization will lie in the out-of-plane (in-plane) direction.
Nevertheless, not all the interstitials are thermodynamically stable, as indicated by the formation energy. 
The candidates with an MAE more than 0.4 MJ/m$^3$ and a negative formation energy are listed in Table~\ref{table1}.



We notice all the parent Heusler compounds listed  in Table~\ref{table1} are ferromagnetic apart from Mn$_2$VGa and Rh$_2$NiSn.  
In our high throughput calculations, for convenience, all Heusler compounds are assumed to be ferromagnetic (FM).   
Previous studies~\cite{RH2MNAL-JPD,RH2MNAL-PRM} have shown Rh$_2$MnAl is an antiferromagnet with Mn are antiferromagnetic coupling between nearest neighbors in the (111) plane,
which is still in the same antiferromagnetic phase after incorporating C or N interstitials.
 As to Mn$_2$VGa, experimental research~\cite{jmm-mn2vga}  has shown it is a half-metallic ferrimagnet with antiferromagnetic coupling between Mn and V with a total net saturation  magnetization  per formula unit as 1.88 $\mu_{\rm B}$ at 5 K. After inducing interstitial (C, B or N),  Mn$_2$VGa is still ferrimagnettic with antiferromagnetic coupling between Mn and V, although initial spin configuration is ferromagnetic.  Mn$_2$VGa have large MAE values as 1.82 MJ/m$^3$, 1.50 MJ/m$^3$ and 1.26 MJ/m$^3$ with B, C and N interstitial, respectively. However, due to the ferrimagnettic phase, the resulting magnetization densities for Mn$_2$VGa with B, C and N interstitial are as weak as about 0.04-0.05 $\mu_{\rm B}$/\AA$^3$. Among all listed compounds in Table~\ref{table1}, Rh$_2$NiSn is  weak ferromagnetic as experimental study~\cite{rh2nisn} suggests it has a magnetic moment 0.6 $\mu_{\rm B}$ per formula unit.  Our calculations demonstrate that H interstitials can induce a tetragonal distortion of c/a=1.26 and a sizable MAE value as 0.82 MJ/m$^3$, whereas the magnetization is only about 0.02 $\mu_{\rm B}$/\AA$^{3}$.

As shown in Table~\ref{table1}, we found 32 compounds with a large out-of-plane MAE (> 0.4 MJ/m$^3$) 
 as well as 10 compounds with large in-plane MAE (absolute value larger than 0.4 MJ/m$^3$). 
In general, the interstitial atoms prefer to be located at the octahedral centers (including both the 24f and 24g sites) except for the H interstitials in Au$_2$MnAl which is stable at the tetrahedral center. 
For the cases of octahedral center, the interstitials mostly prefers 24f sites ($\frac{1}{4}$,0,0) where there are the same atoms in the plane perpendicular to the c-axis.
On the other hand, for Co$_2$FeAl with N, Au$_2$MnAl with N and C, Ni$_2$MnSn with B, C and N, interstitials prefer 24g sites ($\frac{1}{2}$,$\frac{1}{4}$,$\frac{1}{4}$).
We note that Fe$_3$Ge with H  interstitial has  the largest magnetization density as 0.13 $\mu_{\rm B}$ /\AA$^3$ as well as quite large MAE value (1.50 MJ/m$^3$),
 indicating it is a promising permanent magnet. Furthermore, comparing with the magnetization and MAE of experimentally realized permanent magnets~\cite{gutfleisch_magnetic_2011,mae-mag1,mae-mag2,mae-mag3,mae-mag4,mae-mag5}
 Heusler alloys with interstitials  can fill the gap between the low performance magnets (such as AlNiCo and ferrite) and high performance magnets (such as Sm-Co and Nd-Fe-B) in terms of MAE and magnetization, which can spread a wide spectrum of applications.








 
Interestingly, Au$_2$MnAl with H is the only candidate where the interstitials prefer the tetrahedral center (16e site). 
 However, for cases of Au$_2$MnAl with N and and C, interstitial prefers to be located in the octahedral centers with in-plane  MAE. 
  Such special interstitial behaviors can be easily understood based on the chemical bonding.
Intuitively, due to the large atomic spheres of Au atoms,  there is more space between the tetrahedron edge bound than the other Heusler compound. 
For instance, in Au$_2$MnAl with H, the bond length of H-Au pair in  tetrahedral center (1.83 \AA) is comparable with that in octahedral center (1.93 \AA). 
On the other hand, the bond length of Cu-H pair for Cu$_2$MnAl with H in tetrahedral center is just 1.62 \AA,   
 of which the value is obviously smaller than the H-Au pair for H interstitial in the tetrahedral center of Au$_2$MnAl (1.83 \AA). 
 This suggests Au atom can really provide more space for interstitials in the tetrahedral site.  It should be noticed that  Cu-H pair in octahedral center also has a bit larger bond length (1.70 \AA) than that in tetrahedral center. 
 However, in the tetrahedral center case, the bond length is too small to provide enough space for the interstitials. 
  Thus, the H interstitials prefer the octahedral  centers in Cu$_2$MnAl. 
On the other hand, for Au$_2$MnAl with C and N, it is observed that the interstitial atoms still prefer the octahedral center because of the larger atomic radii of  C and N atoms compared to that of H.
Therefore, in order to get the interstitials incorporated at the tetrahedral center, two conditions should be satisfied: (a) The interstitial atoms should be small; (b) There should be large atoms in the parent compound, providing more space. 
Different site preference of the H and C/N interstitials induces significant changes on the MAE of Au$_2$MnAl, {\it e.g.}, H-interstitials favor out-of-plane magnetization while C/N interstitials lead to in-plane magnetization.



\begin{figure}
\centering
\includegraphics[width=6.cm]{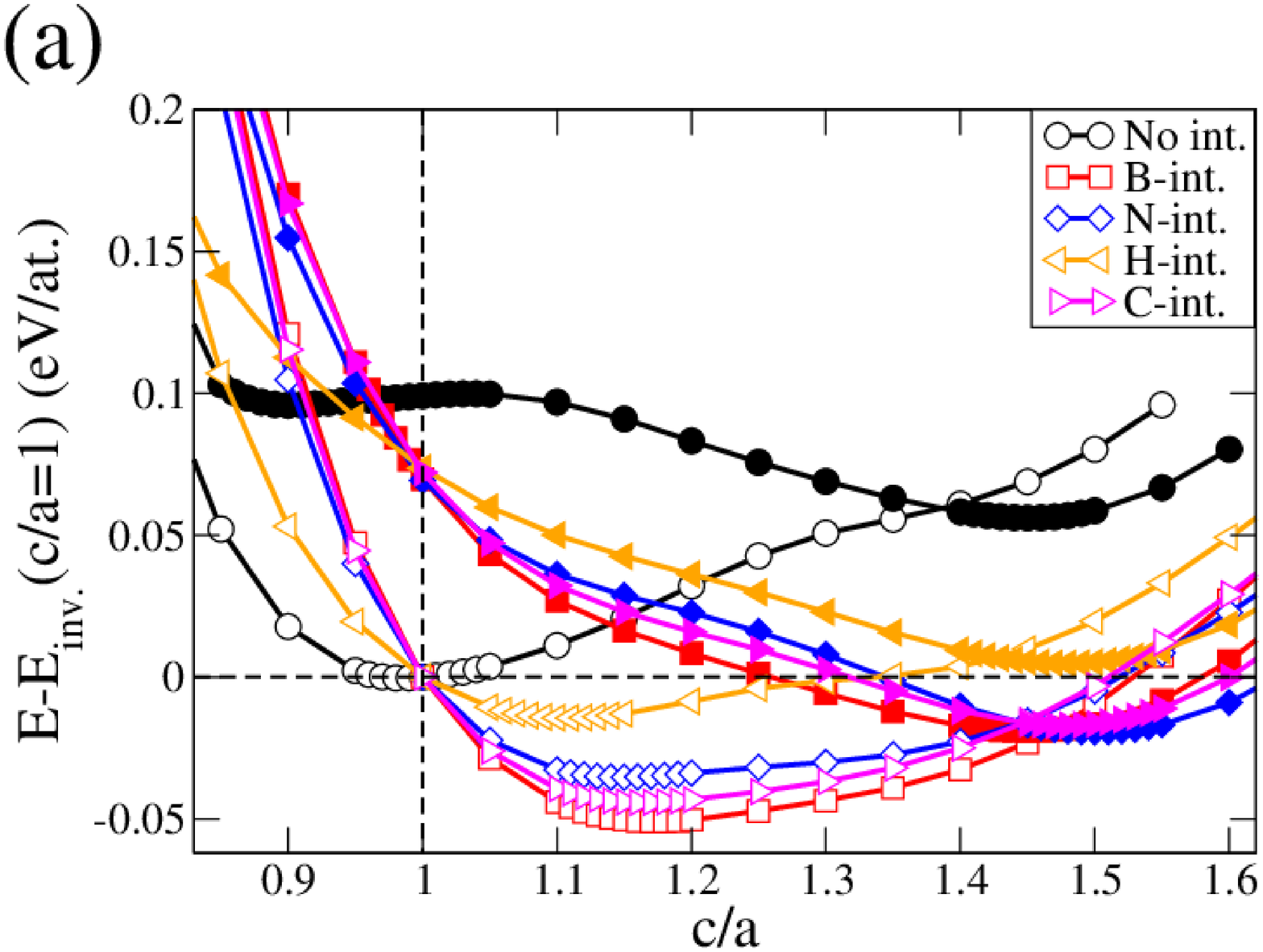}
\includegraphics[width=6cm]{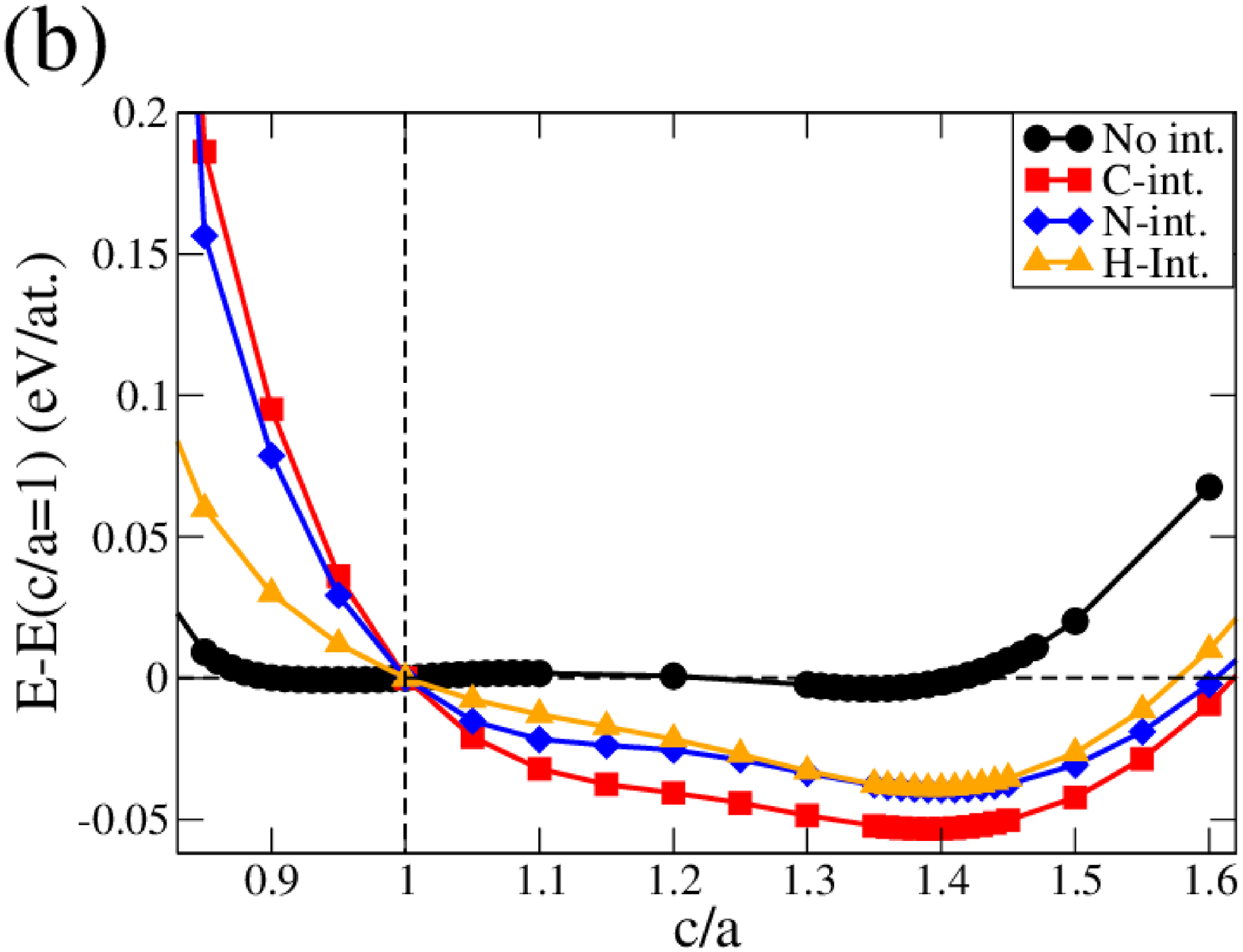}

\caption{}
\label{path}
\end{figure}

According to Table.~\ref{table1}, Fe$_2$CoGa with interstitials is a promising candidate for permanent magnets. However, in the ICSD database~\cite{icsd}, Fe$_2$CoGa (ICSD ID: 102385 and 197615) and Fe$_2$CoGe (ICSD ID: 52954) are in the full Heusler structure, while early M\"ossbauer measurements have shown Fe$_2$CoGa and Fe$_2$CoGe are energetically favored in the inverse Heusler structure~\cite{res1, fe2coge}.
Previous theoretical study~\cite{res2} found that full Heusler Fe$_2$CoGa have 
a martensitic phase transition with a c/a ratio as  1.4, which is also confirmed by our calculation (cf. Fig.~\ref{path}(a)).
 According to our Bain-path calculations,  the inverse Heusler structure is still more energetically favored for Fe$_2$CoGa, even after considering H, B, C, and N interstitials.
Nevertheless, after introducing interstitials, for the full Heusler structure,  the c/a ratio is near to 1.4; whereas for the inverse 
 Heusler structure, the c/a ratio of Fe$_2$CoGa with interstitials is just from 1.1-1.2 due to there is no metastable phase (Fig.~\ref{path}(a)). 
The MAE values of the  inverse Fe$_2$CoGa with  B, C, N, and H  interstitial  are 0.1026 MJ/m$^3$,  0.2148  MJ/m$^3$,  0.3798  MJ/m$^3$, and 0.1925 MJ/m$^3$, respectively. Such lower MAE values are partially due to that interstitials induce much weaker tetragonal distortion to Fe$_2$CoGa for inverse Heusler structure ($1.1\leq c/a \leq 1.2$) than  that for full Heusler structure ($1.45 \leq c/a \leq 1.5$).

More interestingly, C, N, and H interstitials induce significant MAEs to Ni$_2$FeGa. 
 Experimental studies suggest that Ni$_2$FeGa can be grown by melt-spinning technique~\cite{apl-82-3-2003} or glass-purify method~\cite{JournalofCrystalGrowth-388-15-2014},  transforming from high chemical ordering L2$_1$ structure (full Heusler) to martensitic structure at 142 K with a  high Curie temperature of 430 K\cite{apl-82-3-2003}.
Further experiments showed polycrystalline alloys Ni$_{53+x}$Fe$_{20-x}$Ga$_{27}$ have  smaller but comparable  entropy changes as classical magnetocaloric Heusler alloy systems  Ni-Mn-Ga and Ni-Mn-Sn~\cite{apl-132503}. 
DFT calculations suggest that Ni$_2$FeGa has a tetragonal (corresponding to the martensitic phase) structure of c/a=1.35~\cite{jap-111-033905-2012,ni-based} with  an MAE as 0.318 MJ/m$^\text3$~\cite{ni-based}. We also found that Ni$_2$FeGa  is stable in the tetragonal structure with a c/a ratio as 1.35 (Fig.~\ref{path}(b)) and a comparable MAE as 0.2334 MJ/m$^\text3$ (0.0698 meV per chemical formula cell).
 However, the energy difference between tetragonal and cubic structures is as small as 2.80 meV/atom. As proposed by Barman, the martensite phase transition temperature is proportional to the energy difference between cubic and martensite phases~\cite{Barman}, as manifested by the experimental martensitic transition at 142 K~\cite{apl-82-3-2003}.
 After inducing interstitial C, H, or N, Ni$_2$FeGa is stable in the tetragonal phase with $c/a\approx 1.40$.
Correspondingly, the MAEs have been enhanced to 1.43 MJ/m$^3$, 0.94 MJ/m$^3$, and 0.56 MJ/m$^3$ for Heusler Ni$_2$FeGa with N, C, and H interstitials, respectively. 
Obviously, C and N interstitials cause more significant enhancement on the MAE than the H interstitials, though the resulting c/a ratios are comparable.
Therefore, we suspect that both the tetragonal distortion and the chemical bonding environment will influence the MAE values for Heusler with interstitial, which will be discussed in detail below. 



Turning now to the origin of the induced MAE by interstitials, from the theoretical perspective, beside the shape anisotropy due to the magnetic dipole-dipole interaction, 
the magneto-crystalline anisotropy (MCA) can be attributed to the spin-orbit coupling (SOC), which is the dominant contribution to MAE and hence coercivity for PMs.
Based on the perturbation theory, Bruno~\cite{PRB-39-865-1989} pointed out that the MCA can be formulated as
\begin{equation}
\label{eq1}
\text{MCA}=-\sum_i\frac{\xi_i}{4\mu_B}\Delta \mu_{i},
\end{equation}
where $\xi_i$ denotes the atomic SOC constant and $\Delta \mu_{i}$ 
is the orbital moment difference
between the magnetization directions parallel to [001] and [100] for the $i$-th atom. 
We note that such a model is best applicable for strong magnets where the majority spin channel is almost fully occupied, whereas there is
a more general formula considering the spin-flip and quadruple terms~\cite{soc-van}.
Taking Ni$_2$FeGa as an example, Table~\ref{Table:s1} shows the atom-resolved orbital moments and the resulting contributions to the MCA using Bruno's formula, 
where the atomic SOC constants for Ni and Fe are  630 cm$^{-1}$ (corresponding to 78.1100 meV) and  400 cm$^{-1}$ (corresponding to 49.5937 meV)
taken from Ref.~\cite{jcm-soc}. 
The resulting MCA for Ni$_2$FeGa with C, N, and H interstitials based on the Eq.~(\ref{eq1}) are 0.316 meV/f.u., 0.528 meV/f.u. and  0.330 meV/f.u., respectively. Correspondingly,  the  MAEs based on  Eq.~(\ref{eqm}) are 0.296 meV/f.u., 0.439 meV/f.u. and 0.167 meV/f.u., respectively. The relative MAE differences of Bruno's  model to that of Eq.~(\ref{eqm}) are 17.17\%, 23.83\% and 79.61\%.   
Nevertheless, the trend is correctly reproduced and we believe the atomic-resolved contributions evaluated based on Eq.~\ref{eq1} are still valuable to elucidate the origin of MCA.
It is noteworthy that the tetragonal distortion ratios for Ni$_2$FeGa with H, C and N interstitial are 1.39, 1.40 and 1.40, respectively (cf. Table~\ref{table1}).
To make a direct comparison to the pristine Ni$_2$FeGa, we evaluated the MCA and orbital moments for Ni$_2$FeGa without interstitials but with imposed c/a=1.40, resulting in an MCA of 0.066 meV  and  0.170 meV per chemical formula  by using Eq.~(\ref{eqm}) and Bruno's model Eq.~(\ref{eq1}), respectively. 
Again, the MAEs  obtained from the Bruno's model can be well compared with that from Eq.~(\ref{eqm}) for  Ni$_2$FeGa with C and N interstitial, but not for H interstitial case and parent compound. 
 

The remarkable variation of the orbital moments and the resulting significant enhancement of MCAs can be attributed to the atoms surrounding the interstitial atoms.
It is noted that C and N interstitials can give rise a significant MCA to Ni$_2$FeGa, while the effect of H interstitial is rather weaker. 
Following Table~\ref{Table:s1}, it is clear that without interstitials (c/a=1.40), Fe atoms have the leading contribution to the MCA of 0.26 meV per atom, 
while the contribution from Ni (about -0.039 meV per atom) is an order of magnitude lower with opposite sign.  
The change in c/a from 1.35 to 1.40 has minor influence on the MCA and orbit moment.
After considering interstitial N (H), the contribution for Ni-i atoms within the same plane is enhanced to 0.250 meV (0.097 meV) per atom.  
As to C interstitial atoms, the MCA of  Ni-i atoms is slightly (Fig.~\ref{fig:1}(b)) increased to 0.039 meV per atom.
That is, all types of the interstitial atoms lead to a sign change of the contribution to MCA for Ni-i.
On the other hand, the orbital moments and thus the resulting MCA contribution are very comparable for the Ni-ii atoms with and without interstitials,
because the Ni-ii atoms are far away from the interstitials. 
Furthermore, for the H interstitial case, both the MCA and orbital moments of all Fe (including iii, iv and v) atoms change only slightly comparing to those in the pristine compound with imposed c/a=1.40,
whereas the N and C interstitials lead to significant enhancement of contribution for Fe-iii atoms to MCA.
For instance, the MCA contributions of Fe-iii atoms below the interstitials are increased to 0.681 meV and 0.508 meV per atom with N and C interstitials, more than two times larger that (0.260 meV) in the parent compound.
Meanwhile the contributions from Fe-iv and Fe-v atoms are slightly reduced.
Therefore, the interstitial atoms have very strong influence on the MCA of the local surrounding atoms, while the global tetragonal distortion has relatively marginal effects.

The effects of interstitials on MCAs and orbital magnetizations can be further understood based on the chemical
bonding pairs between the interstitials and surrounding magnetic atoms.
For instance, the octahedral center (interstitial) H, C and N to octahedral planar corner Ni-i almost have the same bond lengths as 1.85  \AA,  1.88  \AA \space and 1.88  \AA. However, the integrated COHP of H-Ni  bond is just -0.63 eV, which is much weaker than that of the comparable bond strength of C-(Ni-i) (-2.12 eV) and N-(Ni-i) (-1.96 eV) bonds. 
On the other hand, the octahedral below  corner Fe-iii to the  interstitial H, C and N have similar bond lengths as 1.65 \AA,  1.83 \AA \space and 1.83 \AA, while the bond integrated COHP of H-Fe (iii) (-1.24 eV) can be comparable to that of C-(Fe-iii) (-2.88 eV) and N-(Fe-iii) (-2.38 eV) bonds.   
Obviously, the bond strengths of C interstitial to Fe-iii) and Ni-i are the strongest. This explains the significant change of orbital  moments of Ni$_\text{2}$FeGa with C interstitial comparing to Ni$_\text{2}$FeGa at the same tetragonal distortion ratio without interstitial.   
In this view, the H interstitial just induce tetragonal distortion to Ni$_2$FeGa,  while C and N interstitial not only induce tetragonal distortion but also change the chemical environment by forming strong bonds.
We notice that the interstitial to planer Ni-i atom  have comparable bond lengths as to lower Fe-iii atoms but weaker integrated COHP for each interstitial cases. Such bonding behaviors explains the effect of interstitial on the magnetization and the  MCA for Fe-iii atoms is stronger than that for Ni-i atoms.



\section{Conclusion}

Based on high-throughput DFT calculations, we investigated the effects of (H, B, C, and N) interstitials on the magnetic properties of cubic full Heusler compounds. 
We identified 32 compounds with substantial uniaxial MAE.
Detailed analysis reveals that in addition to the breaking of the cubic symmetry, the changes in the local crystalline environment can induce significant contribution
to the MAE, which can be attributed to the chemical bonding between the interstitial and surrounding magnetic atoms. This could provide an efficient way to
design permanent magnets, which shall be explored further both experimentally and theoretically.


\section*{ACKNOWLEDGMENTS}

Qiang Gao thanks the financial support from the China Scholarship Council. The authors gratefully acknowledge computational time on the Lichtenberg High Performance Supercomputer.

\footnotesize{

}

\newpage

\section*{Captions}

\subsection*{\textbf{Figure captions:}}

\textbf{Figure~\ref{fig:1}: } (a) The possible interstitial sites in the convention Austenite unit cell of full Heusler compounds $X_2$YZ. 
The blue and green octahedrons denote the 24f (0.25,0,0) and 24g (0.5,0.25,0.25) interstitial sites, while the red and pink tetrahedrons mark the 16e(0.125,0.875,0.875) and 16e(0.875,0.625,0.875) interstitial sites.
(b)  
The crystal structure
for the tetragonal full Heusler compound Ni$_2$FeGa with interstitial (int.) at the most stable octahedral sites.

\textbf{Figure~\ref{path}: } Total energy as a function of tetragonal distortion ratio (c/a) for Fe$_2$CoGa with and without interstitials. The reference energy is the energy of the compound in cubic inverse Heusler structure for Fe$_2$CoGa with each interstitial as well as the parent compound. 
 The opened and filled symbols represent the results Fe$_2$CoGa in inverse and full Heusler structures, respectively. 
(b) The total energy as a function of tetragonal distortion ratio for Ni$_2$FeGa in full Heusler structure with and without interstitials.  Here the reference energy is the energy of the compound in cubic full Heusler structure.

\subsection*{\textbf{Table captions:}}

\textbf{Table~\ref{table1}: }   The basic information of the most promising candidates of Heusler compounds with interstitials, where ``site" marks the energetically preferred interstitial site, $\Delta H$ indicate the formation energy in unit of eV/atom, c/a ratio of resulting lattice constants along c-axis and in-plane, MAE in MJ/m$^3$ and meV/f.u. (in parenthesis),  total magnetic moment  M$_\text{tot}$ in the unit of $\mu_{\text B}$/f.u., and the magnetization M/V  in the unit of $\mu_{\rm B}$/\AA$^{3}$. It should be notices the general chemical formula for Heusler compound with interstitial is X$_2$YZI$_{1/4}$, where I is the interstitial.

\textbf{Table~\ref{Table:s1}: } The orbital moment ($\mu_l$, in unit of $\mu_B$) and the magneto-crystalline anisotropy energy (MCA, in unit of meV) energy values for Ni$_2$FeGa with and without interstitials.  Here the MCA is evaluated from Bruno's formula.
 $\mu_l$  and  $\sum$  (in unit of meV) denote the difference of orbital moment between two magnetization directions  ([001] and [100]) and the summation of MCA energy, respectively. The general chemical formula for Heusler Ni$_2$FeGa with interstitial is Ni$_2$FeGaI$_{1/4}$, where I is the interstitial.

\newpage

\section*{Figures}

\newpage

\section*{Tables}

\begin{table}
\caption{}

\label{table1}

\scriptsize
\begin{tabular}{lllllllllllllllllllllll}

\hline
\hline
	Parent & int. & site&$\Delta H$ &c/a & \multicolumn{1}{c}{MAE}& M$_{\text{tot}}$  &M/V \\
	\hline

Fe$_2$CoGa & B&24f& -0.0616  & 1.45    &1.4949 (0.4998)    &5.56 & 0.1120   &   \\
& C& 24f&  -0.0486 & 1.48&  1.3017 (0.4072)   & 5.37 &0.1092  \\
&N &24f & -0.1088 &1.50& 1.3180 (0.4295)    &      5.36&  0.1089 \\
&H &24f& -0.0922  & 1.48&2.3677  (0.6863) &5.95   & 0.1227  \\
\hline

Ni$_2$FeGa& C&24f& -0.1207  &1.40& 0.9636   (0.2961) &2.93 &  0.0595 &   \\
&N &24f & -0.1620 &1.40&1.4292  (0.4386) &2.97   & 0.0602  \\
&H&24f&   -0.1916  &1.39&0.5582 (0.1668)&    3.13  &0.0654  \\
\hline

Fe$_2$CoGe& H&24f& -0.0627 & 1.51&0.6291 (0.2080) &5.48  &0.1142   \\

&N &24f & -0.0576 & 1.56& 0.4047  (0.1368) &5.01  & 0.1025  \\
\hline

Fe$_2$NiAl& H& 24f&-0.2545  & 1.53  &  0.4947 (0.1298) &4.56  &0.0955   \\

&N&24f&-0.2842 &1.57&0.5270 (0.1368) &4.40 &   0.0902\\

\hline

Fe$_2$NiGa
&H  &24f  & -0.1915  & 1.55 &0.5670  (0.1298) &4.67    &0.0977\\
&  B&24f&  -0.1219  &  1.51& 0.7853 (0.1758)  &  4.42&    0.0893            \\
 &C&24f&   -0.1208   &1.53 &  0.9217  (0.1863)  &  4.31& 0.0874 \\
 & N& 24f& -0.1620     &1.53  &  1.3295 (0.2429) &4.45 &0.0939 \\

\hline

Co$_2$MnGa &C&24f&   -0.1289 & 1.13&   0.5267 (0.6303) &  4.52 &0.0920 \\
 &N&24f& -0.1836 &1.12&   0.4755 (0.1576) &4.76 & 0.0922 \\
 \hline

 Co$_2$MnGe&C&24f& -0.0788 & 1.25& 0.5388 (0.1226)  &4.06 &0.0831 \\
&N&  24f& -0.1226 &  1.29&0.5476 (0.1325)  &4.13   &   0.0841     \\

Co$_2$MnSi &C &24f &  -0.2571 &1.21&0.5384  (0.1464) &4.16&0.0898 \\
\hline

 Rh$_2$MnAl &C&24f&-0.5088 &  1.10&   0.9501 (0.3336)   &4.25 &  0.0742  \\

        &N&24f& -0.5457 &1.06&1.1675  (0.4487)  &4.49  & 0.0784  \\

\hline

Rh$_2$NiSn&H& 24f & -0.2288 &  1.26   &  0.8236  (0.3063)  &0.99& 0.0166  \\

\hline

Mn$_2$VGa& C  &  24f  &-0.1533 &  1.20&	1.5038  (0.4874) &	2.26&0.0435    \\
&B &24f & -0.1474 & 1.23&  1.8263  (0.5987) & 2.48&   0.0472        \\
&N& 24f& -0.2377 & 1.21& 1.2674  (0.4087) &  2.34  &		0.0451			\\
\hline

Co$_2$FeAl&N&24g& -0.2770 & 1.08&  0.4881  (0.3009)  &5.01&0.1038   \\

\hline

Au$_2$MnAl & H&16e&-0.1835 & 0.92&0.7732  (0.6489) & 3.82&0.0582  \\
&N & 24g&-0.1975 & 1.27 & -0.4091 (-0.2412)&3.66 & 0.0476  \\
 &C& 24g &-0.2770  & 1.21& -0.5271 (-0.4923)  &3.82 &0.0574\\
\hline

Ni$_2$MnIn&C &24f&  -0.0057 & 1.21&  -1.0288 (-0.3513) &  3.96 & 0.0685\\

\hline
Ni$_2$MnGa&H& 24f &-0.2519  & 1.27&-1.3278  (-0.4898) &4.20  &0.0852&   \\

 &  B& 24f&  -0.2204    & 1.28 &-0.5822 (-0.1850) & 4.02&  0.0788  \\
& C &24f& -0.1780 & 1.29&  -0.9573 (-0.3031) & 3.91& 0.0771  \\

\hline

Fe$_3$Ge&H& 24f &-0.0563  & 1.42&1.5018 (0.4655)  &6.44   & 0.1319  \\
&  B &   24f& -0.0254     &   1.16&  0.5868 (0.1812) &  5.54 &    0.1211         \\
\hline

Fe$_3$Ga& B&24f&  -0.0710 &            1.21&0.6896  (0.2142) & 6.16&  0.1229 \\
&N& 24f& -0.1101 &     1.19&  0.5184 (0.1610) &  5.92  &0.1022   \\
\hline

Ni$_2$MnSn& B& 24g& -0.0959   &1.06& -0.6747  (-0.2064) &3.75& 0.0645 \\
&C& 24g& -0.0480 &  1.17&-0.4261 (-0.1421)&  3.70 & 0.0653\\
&N&24g &     -0.0892   &  1.17& -0.4756 (-0.1613) &  3.74 &  0.0697  \\
\hline
  Rh$_2$MnSn&  C& 24f&  -0.2679 & 1.26 &    -0.8846 (-0.3529) &  3.66 &0.0572             \\


\hline
\hline

\end{tabular}
\end{table}

\begin{table*}[]
\centering
\caption{}
\begin{tabular}{|l|l|l|l|l|l|l|l|l|}

\hline
                                & int.                &         & Ni-i & Ni-ii & Fe-iii & Fe-iv & Fe-v & $\sum$ \\ \hline
\multirow{12}{*}{\rotatebox{90}{ with int.}} & \multirow{4}{*}{H} & [001] &  0.024&0.020 & 0.068   &   0.063  & 0.066      &   \_    \\ \cline{3-9} 
                                 &                    & [100] &   0.019& 0.022   &  0.043   & 0.040   &  0.046 &   \_    \\ \cline{3-9} 
                                 &                    & $\Delta$  & 0.005  & -0.002   & 0.025    &   0.023 &  0.020 &   \_    \\ \cline{3-9} 
                           &                    & MCA    &   0.097&  -0.039  & 0.309    & 0.284   & 0.247  &    0.330    \\ \cline{2-9}

                           & \multirow{4}{*}{N} & [001] &  0.025 &  0.020  & 0.068    & 0.067   &  0.072 &    \_    \\ \cline{3-9} 
                           &                    & [100] & 0.012  &0.024    & 0.013    &  0.053  & 0.049  &    \_    \\ \cline{3-9} 
                           &                    &$\Delta$  & 0.013  &  -0.004  &   0.055  &  0.014  &  0.023 &    \_    \\ \cline{3-9} 
                           &                    & MCA    &  0.250 &-0.078    &  0.681   &  0.173  &0.285   & 0.528      \\ \cline{2-9}

                           & \multirow{4}{*}{C} & [001] & 0.013  &  0.022  &    0.058 &  0.067  & 0.073  &   \_     \\ \cline{3-9} 
                           &                    & [100] & 0.011  &0.024    & 0.017    &0.048    &  0.051 &   \_     \\ \cline{3-9} 
                           &                    & $\Delta$ & 0.002 & -0.002  &  0.041  &  0.019   &   0.021 &    \_       \\ \cline{3-9} 
                           &                    & MCA     & 0.039  & -0.039   &  0.508   & 0.235   &  0.260 &   0.316    \\ \hline

                          & c/a                   &   &  \multicolumn{2}{l|}{Ni} & \multicolumn{3}{l|}{Fe} &  $\sum$ \\
                          \hline
\multirow{8}{*}{\rotatebox{90}{w/o int.} }& \multirow{4}{*}{1.35} & [001]   & \multicolumn{2}{l|}{0.022}   & \multicolumn{3}{l|}{0.065}   &-  \\
 \cline{3-9}
                          &                       & [100]  &\multicolumn{2}{l|}{0.024}   & \multicolumn{3}{l|}{0.044}   &-  \\
                           \cline{3-9}
                          &                       &   $\Delta$ & \multicolumn{2}{l|}{-0.002}   & \multicolumn{3}{l|}{0.021}   &  - \\
                          \cline{3-9}
                          &                       &  MCA  &\multicolumn{2}{l|}{-0.039}   & \multicolumn{3}{l|}{0.260}   & 0.180  \\

 \cline{2-9}                          
                          & \multirow{4}{*}{1.40}     &  [001] &\multicolumn{2}{l|}{0.021}   & \multicolumn{3}{l|}{0.061}   &-   \\
   \cline{3-9}                       
                          &                       & [100]   & \multicolumn{2}{l|}{0.023}   & \multicolumn{3}{l|}{0.041}   & -  \\
      \cline{3-9}                    
                          
                          &                       &  $\Delta$    & \multicolumn{2}{l|}{-0.002}   & \multicolumn{3}{l|}{0.020}   &   \\
                          \cline{3-9}
                          &                       & MCA   &\multicolumn{2}{l|}{-0.039}   & \multicolumn{3}{l|}{0.248}   &  0.170

\\
\hline
\end{tabular}
\label{Table:s1}

\end{table*}

\newpage

\section*{Highlights}

\begin{itemize}

\item Rare earth free permanent magnets can be realized in tetragonally distorted full Heusler alloys induced by light interstitial atoms.

\item Bain path calculations reveal that interstitials cause stable tetragonal distortion to full Heusler alloys.

\item Analysis based on the perturbation theory and chemical bonding suggests that the uniaxial anisotropy can be attributed to change in the local crystalline environments around the interstitials.

\item We postulate that this provides a universal way to tailor the magnetic properties of prospective permanent magnets.

\end{itemize}

\newpage

\appendix

\section{Appendix}
\label{appendx}



\renewcommand{\thetable}{A\arabic{table}}

\begin{table*}
 \renewcommand\thetable{A.1}
\caption{All the considered Heusler compounds (Com.) together with the ICSD ID number.}

\label{tables1}

\footnotesize
\begin{tabular}{|l|l||l|l||l|l||l|l|lll}
\hline
\hline
	Com. & ID &Com. & ID &Com.&ID &Com.&ID\\
	\hline
	  Au$_2$MnAl& 57504&Co$_2$CrAl&57600& Co$_2$FeAl& 57607& Co$_2$HfAl&   110809       \\
	\hline
Co$_2$MnAl&   606611&Co$_2$NbAl&57620 & Co$_2$TaAl&  606667 &  Co$_2$TiAl &   606680         \\
\hline

    Co$_2$VAl &   57643 &    Co$_2$ZrAl & 57648&  Co$_2$CrGa &   102318&   Co$_2$CrIn&  416260      \\
    \hline
Co$_2$FeGa&102392   &Co$_2$FeGe&  247268&  Co$_2$FeIn& 102392  &  Co$_2$FeSi& 622985\\
\hline

Co$_2$HfGa& 102433  & Co$_2$MnGa& 623116&Co$_2$NbGa& 623126& Co$_2$TaGa &102451      \\	
\hline

Co$_2$TiGa   &  102453& Co$_2$VGa   &  623228  &   Co$_2$LiGe  & 53673  &  Co$_2$MnGe  & 52971  \\

\hline
Co$_2$TiGe  &  169469  & Co$_2$ZnGe    &    52994&    Co$_2$HfSn   &  102483  &  Co$_2$MnSb  &   53002    \\

\hline
Co$_2$MnSi  &   106484  &   Co$_2$MnSn  &  102332  & Co$_2$NbSn    & 102554  & Co$_2$ScSn   & 102646   \\
\hline

Co$_2$TiSi&   53080  &Co$_2$VSi&    53086  &Co$_2$TiSn&   102583     &Co$_2$VSn&   102684\\
 \hline
 
 	Co$_2$ZrSn   &		102687	 &  	Cu$_2$CrAl  & 57653& 	Cu$_2$MnAl	   &  	607012	&Cu$_2$CoSn	   &	103057	\\
 \hline
 
 Cu$_2$FeSn	   &      151205    &Cu$_2$MnIn& 102996&	Cu$_2$MnSb&     53312 &Cu$_2$MnSn&   103057             \\
 \hline
  
Cu$_2$NiSn&  103069    & Fe$_2$CrAl  &184446  &   Fe$_2$MnAl  & 57806  &  Fe$_2$MoAl &  57807 \\

\hline

Fe$_2$NiAl &    57808  &  Fe$_2$TiAl &   57827 &  Fe$_2$VAl &  57832   &  Fe$_2$CoGa&   103473\\

\hline

Fe$_2$CoGe&  52954  &Fe$_2$CrGa  &   102755  & Fe$_2$NiGa   &103460   &  Fe$_2$TiGa   &  103469
\\
\hline

 Fe$_2$VGa   &  103473   & Fe$_2$MnSi   &    632569  &  Fe$_2$VSi   &   53555&    Fe$_2$TiSn&    103641\\
 
\hline

Fe$_2$VSn&   103644  &  Mn$_2$VAl &   57994   &  Mn$_2$RhGa   &    247951   &  Mn$_2$VGa   &103813   \\
\hline

 Mn$_2$RuGe  &   247950  &  Mn$_2$RuSn   &  247949   &  Mn$_2$WSn   &  104980 &   Ni$_2$CrAl  &  57662     \\ 

\hline
Ni$_2$HfAl  & 57901   & Ni$_2$MnAl  & 57976  &   Ni$_2$NbAl  &  58016  &  Ni$_2$ScAl  &   58050    
\\ \hline

 Ni$_2$TaAl  &    58055  &   Ni$_2$TiAl  &   58063    &   Ni$_2$VAl  & 58071 &   Ni$_2$ZrAl  &  58081\\
 \hline
 
 Ni$_2$CuSb  &   53320  &   Ni$_2$CuSn &   103068   &  Ni$_2$HfGa  &  103734  &   Ni$_2$MnGa  &   103803\\
 
 \hline
 
  Ni$_2$NbGa  & 103839  &    Ni$_2$ScGa  &  103874 &    Ni$_2$TaGa  &   103881  &     Ni$_2$TiGa  &   103886\\
  
  \hline
  
   Ni$_2$VGa  &  103892 &    Ni$_2$ZrGa  &  103902  &  Ni$_2$LiGe &  53673  &   Ni$_2$MnGe &  192566\\
   
   \hline
    Ni$_2$ZnGe &   53865  &  Ni$_2$HfIn   &   54595   &  Ni$_2$HfSn   & 104250&  Ni$_2$MgIn   &51982\\
    \hline
 Ni$_2$MnIn   &   639954  &   Ni$_2$ScIn  &  59446  &  Ni$_2$TiIn  &   59451&  Ni$_2$ZrIn  &   59460
 \\  \hline  
  
 Ni$_2$LiSi  &  44819  &    Ni$_2$LiSn  &  25325   &    Ni$_2$MgSb &  104841&    Ni$_2$MgSn &   104842 \\
 \hline
 
   Ni$_2$TiSb &  76700   &   Ni$_2$ZrSb &   76703  &   Ni$_2$ScSn &   105339  &  Ni$_2$TiSn &   105369  \\
\hline

 Ni$_2$VSn &  105376  &   Ni$_2$ZrSn &    105383   &   Pd$_2$MnAl &    57981  &    Pd$_2$MnAs &  107955
 \\ \hline

   Ni$_2$NbSn &   105181  &   Pd$_2$MnGe &   53705 &   Rh$_2$NiSn &  105327   &    Pd$_2$MnIn &  51990  \\
\hline
Pd$_2$MnSb &     643312    & Pd$_2$MnSn &    104945  &  Rh$_2$MnAl &    57986  &  Rh$_2$MnGe &   53706 \\
\hline

 Rh$_2$MnPb &    104936&    Rh$_2$MnSn &  104964  &       Ru$_2$FeSi&  53525  &Ru$_2$FeSn&    103615          \\

 \hline
 
 Fe$_3$Al  &  57793  &   Fe$_3$Ga  &  108436  &  Fe$_3$Ge  &   53462  & Fe$_3$Si  &   53545	         \\ \hline
 
  Mn$_3$Si  &   76227   &   Ni$_3$Al  &      58038    &   Ni$_3$Sb &   76693   &  Ni$_3$Sn &105354
  \\
  \hline\hline

\end{tabular}

\end{table*}

\end{document}